\documentclass[preprint]{revtex4-2}
\usepackage[utf8]{inputenc}

\usepackage{amsmath}
\usepackage{tikz}
\usetikzlibrary{calc}
\usepackage{anyfontsize}
\usepackage{float}
\usepackage{comment}
\usepackage{lineno}

\begin{document}

\title{Emulating ultrafast dissipative quantum dynamics with deep neural networks}
\author{Nikolai D. Klimkin}
\email{nd.klimkin@physics.msu.ru}
\affiliation{Russian Quantum Center, Bolshoy Bulvar 30, bld. 1, Moscow, Russia 121205}
\affiliation{Faculty of Physics, Lomonosov Moscow State University, Leninskie Gory 1-2, Moscow 119991}

\begin{abstract}
    The simulation of driven dissipative quantum dynamics is often prohibitively computation-intensive, especially when it is calculated for various shapes of the driving field. We engineer a new feature space for representing the field and demonstrate that a deep neural network can be trained to emulate these dynamics by mapping this representation directly to the target observables. We demonstrate that with this approach, the system response can be retrieved many orders of magnitude faster. We verify the validity of our approach using the example of finite transverse Ising model irradiated with few-cycle magnetic pulses interacting with a Markovian environment. We show that our approach is sufficiently generalizable and robust to reproduce responses to pulses outside the training set.
\end{abstract}

\maketitle




\section{Introduction}

The simulation of dissipative quantum dynamics \cite{breuer2002theory} is a problem which arises in diverse areas of physics, including ultrafast spectroscopy, chemical physics, quantum optics, quantum biology, quantum computing, and quantum information technology \cite{mukamel1999principles, may2008charge, nitzan2006chemical, nielsen2002quantum, de2017dynamics}. Some problems involving open quantum systems in the presence of a driving field, which include optimal control of open quantum systems~\cite{schmidt2011optimal, abdelhafez2019gradient} and simulating coherent pulse propagation using the Maxwell-Schr\"odinger equations~\cite{lorin2007numerical} when accounting for divergence- and propagation-induced decoherence~\cite{kilen2020propagation, floss2018ab, Wikmark4779}, require simulating the system iteratively for different waveforms of the field. Many methods for simulating open quantum systems are available, including Monte Carlo methods~\cite{stockburger1999stochastic}, time evolving density matrix methods, and path integral approaches. However, they often become unsuitable for iterative methods by making any computations involving them undesirably expensive, and, for optimal control problems, by not having a straightforward way of computing gradients.

Deep neural networks have recently been gaining prominence in physics~\cite{Carleo:19}. They provide a robust and versatile toolset for e.g. regression problems, already being used for such diverse applications as boosting the signal-to-noise ratio in LHC collision data~\cite{Baldi:14}, establishing a fast mapping between galaxy and dark matter distribution~\cite{Zhang:19}, and constructing efficient representations of many-body quantum states~\cite{Sharir:20}. In this work, by introducing a new feature space specifically designed for dissipative dynamics, we reformulate the problem of recovering dynamics as a regression problem. We thus propose a feed-forward neural network-based approach in which we train our network to emulate the strong-field dynamics of a dissipative quantum system by mapping the incident field directly to the target observables. 

Existing approaches to simulating dissipative quantum dynamics with neural networks focused on reconstructing its dynamics using a limited set of physical observations~\cite{flurin2020using, herrera2021convolutional}. Existing machine learning approaches to many-body systems involve constructing an efficient representation of the quantum system's state using restricted Boltzmann machines~\cite{schuld2019neural}. 

By introducing a novel way of representing the incident field and inferring the system observables directly from this representation, our approach becomes agnostic of the system's state, and thus avoids any integration of the system's state over time. This offers a distinct performance advantage over exact integration (see Discussion). At the same time, unlike Kerr nonlinearity-based schemes used for simulating propagation of strong pulses~\cite{chiron1999numerical}, our approach allows to emulate arbitrarily nonlinear processes in the medium, lending itself well to strong-field optics. 

In this work, we consider an example of a quantum system driven by a strong field. As our quantum system, we choose the transverse Ising model with periodic boundary conditions. We use it to demonstrate that a neural network can learn the dynamics of a system which is correlated, not solvable analytically in the presence of a driving field, and has a Hilbert space dimension greater than the hidden spaces generated by the neural network.

\section{Methods}

Suppose a quantum system characterized by a set of observables $O_1,\ldots, O_m$ is initialized in a state $\rho$. Suppose that this state is perturbed by a time-dependent field $F(t)$ and evolves according to a general-form master equation:

\begin{equation}
    \dot\rho = \Phi(F(t))[\rho]
\end{equation}

In this case, the observables can be inferred given complete knowledge of the history of the field before the time point t. Suppose that the system interacts with a Markovian enviroment, and thus has a finite coherence time $T_2$. We hypothesize that the observables can be inferred from the expansion coefficients over the Laguerre polynomial basis scaled by a factor of $T_2$:

\begin{equation}
    O_i(t) = f_i\left(\left\{\int\limits_0^{+\infty} d\tau L_n(2\tau/T_2) e^{-2\tau/T_2} F(t-\tau), n=\overline{1, N}\right\} \right)
\end{equation}

Here we take advantage of the fact that Laguerre polynomials are orthogonal with exponential weight:

\begin{equation}
    \int\limits_0^{+\infty} dx L_m(x) e^{-x} L_n(x) = \delta_{mn}
\end{equation}

Or, alternatively, the Laguerre functions $l_n(x) = L_n(x)e^{-x/2}$ are orthogonal with unity weight. Effectively, we introduce the decaying field $F_d(t; \tau) \equiv F(t-\tau) e^{-\tau/T_2}$, where the decaying exponent accounts for the system losing coherence and memory of its previous state over time. We expand this decaying field over an orthogonal basis of modified Laguerre polynomials $L_n(2\tau/T_2) e^{-\tau/T_2}$, forming a "history vector" $\mathbf{h}(t)$.  

The vector function $\mathbf{f}$ is then parameterized by a deep fully connected neural network. We thus train it to receive the input $\mathbf{h}(t)$ and yield the correct observables $\mathbf{O}(t)$.

The system we pick as example of an abstract quantum model is the 8-site transverse Ising model in the ferromagnetic phase, irradiated by a magnetic pulse directed along the z-axis with a waveform $F(t)$. Using the Peierls substitution, we arrive at the following system Hamiltonian:

\begin{equation}
    \hat{H} = J\sum\limits_{<ij>} \sigma_z^i\sigma_z^j + g\left[\cos A(t) \sum\limits_i \sigma_x^i - \sin A(t) \sum\limits_i \sigma_y^i \right]
\end{equation}

Its only observable that we want to reproduce is the current, which is in the form of:

\begin{equation}
    \hat{O}_1 = \hat{j} = \frac{\delta H}{\delta A} = -g\left[\sin A(t) \sum\limits_i \sigma_x^i + \cos A(t) \sum\limits_i \sigma_y^i \right]
\end{equation}

The coherence decay is introduced using the Lindblad equation~\cite{lindblad1976generators} with the jump operator being in the form~\cite{Weisbrich_2018}:

\begin{align}
    \dot\rho =& i[H, \rho] + \gamma\left(L\rho L^\dag - \frac{1}{2}\{L^\dag L, \rho\}\right)\\
    &\text{ where } \gamma \equiv T_2^{-1}, L = \sum\limits_{E_\alpha<E_\beta} \hat\Pi(E_\alpha) \left(\sum\limits_i \sigma_z^i\right) \hat\Pi(E_\beta)
\end{align}

The above equation is solved for 704 short pulses. The pulses are generated in the following form:

\begin{align}
    F(t) = (A\omega)\exp(-(t')^2/2\sigma^2)\cos(\omega t + \varphi)&, &\text{where } t' \equiv \max(t, 0, \mu - t)
\end{align}
\begin{equation}
    A(t) = \int\limits_{-\infty}^t F(t') dt' 
\end{equation}

The system parameters are set to:

\begin{equation}
    J=-2.4, g=1.0
\end{equation}
\begin{equation}
    A = 1.0\div 16.0, \omega = 0.1\div 0.3, \varphi = 0\div 2\pi, \mu = 0\div 6\pi/\omega
\end{equation}

The observables are recorded and history vectors truncated at $N=10$ are calculated for each time point. After this, the neural network is trained to reconstruct the observables from history vectors.

\begin{figure}
    \includegraphics[width=\textwidth]{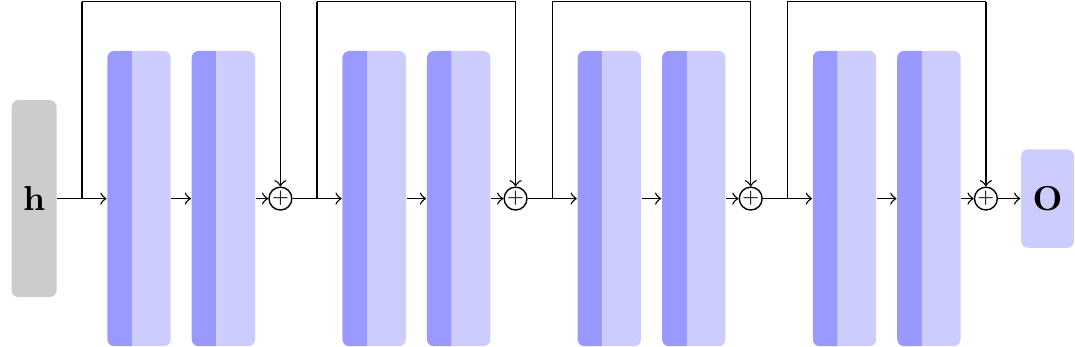}
    \caption{The neural network architecture used in the work. The grey block is the input history vector, the dark blue blocks denote a batch normalization layer with subsequent Mish activation, and the light blue blocks fully connected layers. The neural network is arranged as a sequence of residual blocks which apply a dimension-preserving nonlinear transform $\mathcal{T}$ to the input $\mathbf{x}$, and then construct their output as $\mathcal{O}(\mathbf{x}) = \mathcal{T}(\mathbf{x}) + \mathbf{x}$.}
    \label{fig:nn_arc}
\end{figure}

The neural network, depicted on figure \ref{fig:nn_arc} consists of 4 residual blocks, each consisting of 2 consecutive fully connected layers, with 64 neurons in each, each preceded by a batch normalization layer~\cite{ioffe2015batch} and a Mish activation~\cite{misra2019mish}. It is trained for 2000 epochs with the AdaBelief optimizer~\cite{zhuang2020adabelief} with learning rate 1.0, discounted by a factor of 10 after 1500 epochs. This demonstrates that the neural network can emulate a quantum system while having fewer neurons in the final layers (i.e. fewer dimensions of the hidden space) than either the dimension of the Hilbert space (256) or the number of independent parameters in the density matrix (65534), which means this approach can be readily generalized to larger quantum systems. 

\section{Results}

\begin{figure}[t]
    
    \resizebox{\textwidth}{!}{
    \begin{tikzpicture}[font=\fontsize{48}{48}\selectfont]
    \node[inner sep=0] (pic1) at (0,0) {\includegraphics{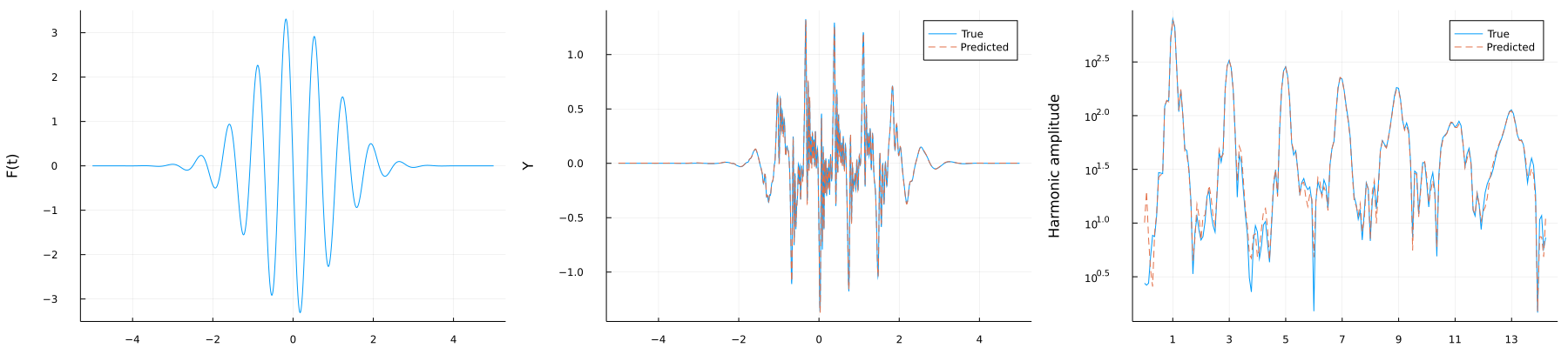}};
    \node[inner sep=0, anchor=north] (pic2) at (pic1.south) {\includegraphics{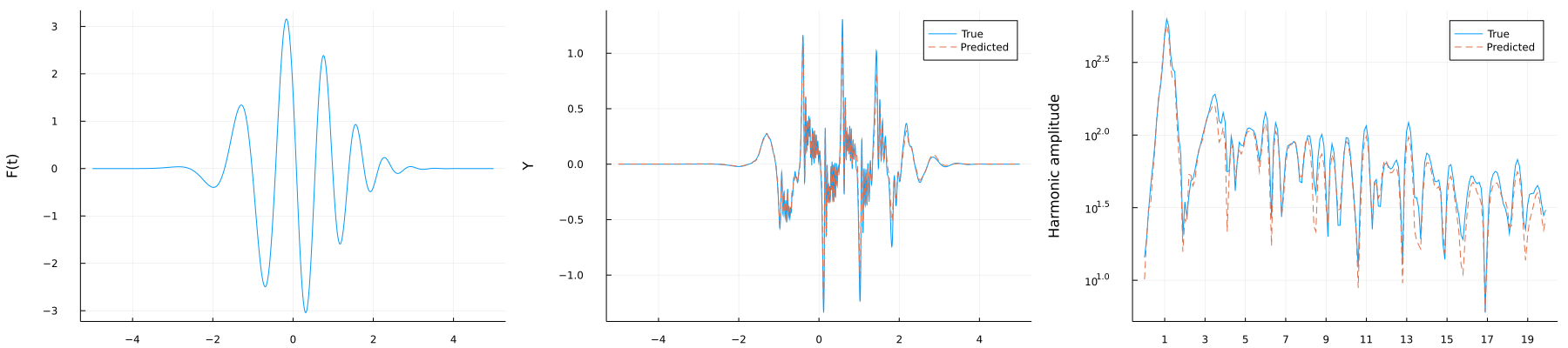}};
    
    \path let \p1=(pic1.north west),\p2=(pic2.south east), \n1={\x2-\x1}, \n2={\y2-\y1} in 
    node[xshift=\n1/10, yshift=\n2/20] (ta) at (pic1.north west) {a)}
    node[xshift=\n1/3] (tb) at (ta) {b)}
    node[xshift=\n1/3] (tc) at (tb) {c)}
    node[yshift=\n2/2] (td) at (ta) {d)}
    node[xshift=\n1/3] (te) at (td) {e)}
    node[xshift=\n1/3] (tf) at (te) {f)};
    
    \end{tikzpicture}}
    
    \caption{Typical emulation results. The top row (a-c) shows the recovery for an unchirped pulse from the test set, the bottom (d-f) for a chirped one, which is not contained in either the test or the training set. The left column shows the incident field $F(t)$, the center the respective generated current $j(t)$, compared to the NN reconstruction, the right the spectrum $|j(\omega)|$, shown between the 0th and the 20th harmonics of the fundamental laser frequency.}
    \label{fig:rec_plots}
\end{figure}

The typical recovery result, taken for high values of $A$ and $\omega$, is demonstrated in Fig.~\ref{fig:rec_plots}~(a-c). We observe that the recovery of the instantaneous values of the dipole moment, as well as frequency amplitudes between H0-H20, are nearly exact. This demonstrates the advantage of our method over approximate models based on the Kerr nonlinearity~\cite{chiron1999numerical}: the simulated dynamics are arbitrarily nonlinear with respect to the field, and not limited to the 3rd harmonic. 

To evaluate the generalization capability of our model, we present it with a qualitatively new example. The model is input the history vectors for a chirped pulse defined by \eqref{eqn:f_chp}, \eqref{eqn:a_chp}, while the training set only contains spectrally-limited pulses. 

\begin{equation}\label{eqn:f_chp}
    F(t) = (A\omega)\exp(-t^2/2\sigma^2)\cos(\omega t + \alpha t^2 + \varphi)
\end{equation}
\begin{equation}\label{eqn:a_chp}
    A=16.0,~\omega = 0.2,~\alpha = \omega^2/10\pi
\end{equation}

We observe on Fig.~\ref{fig:rec_plots}~(d-f) that the accuracy with which the response is recovered is worse, albeit still high across the spectrum.

\section{Discussion}

Our approach offers new pathways for accelerating and optimizing computations involving dissipative quantum systems which may otherwise be undesirably computation-intensive. Thanks to being state-agnostic, our method promises an overwhelming performance advantage over both conventional many-body neural simulators and exact TDSE computations. While an exact Lindblad simulation of the system dynamics over 10 laser cycles requires 12.0 s on average with an NVidia Tesla P100 GPU, its emulation with a trained neural network for all time points at once only takes 1.2 ms. This figure does not include the computation of Laguerre expansion coefficients, which takes an additional 2.9 ms per trajectory in its ''naive'' implementation; however, this can be accelerated by introducing Fourier~\cite{shen2001orthogonal} or transport equation~\cite{terekhov2021generating}-based methods. Although the 8-site Ising model was chosen as example, it can be easily substituted by a different system, and a new dataset can be generated following the same procedure.

A further perspective of our work would be to validate our approach on an iterative problem which was indeed the objective. Most likely, such validation will be done on a dissipative Maxwell+TDSE propagation problem. This may be done in two ways: \textbf{(A)} A Maxwell solver where the polarization term is parameterized by a neural network could be constructed. The neural network will then be trained to reproduce the correct propagated pulses when serving as the right-hand term, rather than the correct dipole moments. This will allow to combat errors which may otherwise accumulate, however small they initially be. \textbf{(B)} The training set could be enhanced to include pulses and responses generated by a TDSE+Maxwell solver during exactly-computed propagation. Preparing such a dataset is more expensive in terms of time and resources, and this approach does not automatically protect one against accumulating errors, but training the NN itself would be easier.

Both of these approaches are to pursue a common goal, namely, to construct a robust simulation where the quantum response of the medium is emulated by a neural network. Although the exact limits of such a simulation, as well as its generalizability to higher dimensions, are yet to be determined, our work demonstrates that such emulation is possible within a broad range of parameters.

\section{Acknowledgements}

We thank Evgeny A. Polyakov, Misha Ivanov, and Alexey N. Rubtsov for fruitful discussions and useful feedback. 

N.K. acknowledges funding by the Foundation for Assistance to Small Innovative Enterprises (agreement No 196GUTsES8-D3/56338), Foundation for the Advancement of Theoretical Physics and Mathematics (agreement No 20-2-2-39-1), and Non-commercial Foundation for the Advancement of Science and Education INTELLECT (agreement No ASP-32-NS\_FF/2021). 

\section{Data availability}

The data supporting the figures are available from the author upon reasonable request.

\bibliography{refs}

\end{document}